\documentclass{article}

\usepackage{graphicx}
\usepackage{amsmath}
\usepackage{amssymb}
\usepackage{eucal}
\usepackage{natbib}

\newcommand{\ocaml}{\texttt{ocaml}}
\newcommand{\nbar}{\ensuremath{\bar{N}}}
\newcommand{\varn}{\ensuremath{\sigma_N^2}}
\newcommand{\Ic}{\ensuremath{I_c}}
\newcommand{\Qone}{\ensuremath{Q_1}}
\newcommand{\Itwo}{\ensuremath{I_2}}
\newcommand{\Bone}{\ensuremath{B_1}}
\newcommand{\Btwo}{\ensuremath{B_2}}
\newcommand{\Aone}{\ensuremath{A_1}}
\newcommand{\Atwo}{\ensuremath{A_2}}

\newcommand{\x}{\mathbf{x}}
\newcommand{\one}{\ensuremath{\mathbf{1}}}
\newcommand{\argmax}{\operatornamewithlimits{argmax}}
\newcommand{\norm}[1]{\ensuremath{\|#1\|}}
\newcommand{\stack}[2]{\begin{smallmatrix} #1 \\ #2 \end{smallmatrix}}

\newcommand{\SBsection}[1]{\vspace{.6cm} \noindent \textsc{#1} \vspace{.2cm}}
\newcommand{\SBsubsection}[1]{\vspace{.4cm} \noindent \textit{#1} \vspace{.2cm}}
\newcommand{\SBsubsubsection}[1]{\vspace{.4cm} \noindent
  \textsc{\small #1} \vspace{.2cm}}

\title{A geometric approach to tree shape statistics}
\author{Frederick A. Matsen}

\begin{document}

\maketitle
 
\newcounter{count}

\begin{abstract}
  This article presents a new way to understand the descriptive
  ability of tree shape statistics. Where before tree shape statistics
  were chosen by their ability to distinguish between
  macroevolutionary models, the ``resolution'' presented in this
  paper quantifies the ability of a statistic to differentiate between
  similar and different trees. We term this a ``geometric'' approach
  to differentiate it from the model-based approach previously
  explored. A distinct advantage of this perspective is that it allows
  evaluation of multiple tree shape statistics describing different
  aspects of tree shape. After developing the methodology, it is
  applied here to make specific recommendations for a suite of three
  statistics which will hopefully prove useful in applications. The
  article ends with an application of the tree shape statistics to
  clarify the impact of omission of taxa on tree shape.
\end{abstract}

The analysis of phylogenetic tree shape provides one way of
understanding the forces guiding macroevolution, as well as
understanding possible biases of tree reconstruction methodology.
Although it has been a subject of study for many years, a recent
editorial in this journal \citep{simon-page} hints that finding the
forces guiding tree shape is a long-term challenge which has yet to be
completely understood. Joe \citet{felsenstein} concludes the chapter
on tree shape methodology in his recent book with the simple phrase
``[c]learly this literature is in its early days.'' Indeed, tree shape
is still a challenge, and an important one. A complete understanding
would help resolve important questions in biology such as the roles of
adaptive radiation and environmental change in generating diversity.
Tree shape also poses difficult issues of its own, such as the impact
of missing or extinct taxa on our understanding of historical
biodiversity. Not only are many fundamental questions left unanswered,
but the area is ripe for progress: the large number and size of
contemporary phylogenies forms a fantastic corpus on which
macroevolutionary hypotheses can be tested.

In order to use phylogenetic tree shape as a tool, we need methods to
measure and quantify aspects of tree shape. Almost all work to this
day has been done with measures of tree ``balance,'' which is the
degree to which two sister taxa are of the same or different size. A
major vein of research has been to compare balance of trees created
from data to trees produced by one or another null model
\citep{Savage1983:225} \citep{Guyer1991:340} \citep{Guyer1993:253}
\citep{stam02a}. \citet{Kirkpatrick1993:1171}, in one of the early
papers in the area, quantified the power of different measures of tree
balance in distinguishing between two models of tree shape. The two
models are extremely simple: one, called the Yule or ERM model,
develops a tree by starting with a single species and then choosing
uniformly among species to speciate. The other, called the PDA model,
is simply the distribution on tree shapes induced by the uniform
distribution on labelled trees.

Studies have shown that most trees created from data are less balanced
than would be expected from the ERM model, yet more balanced than
would be expected from the PDA model \citep{Mooers1995:379}
\citep{Mooers1997:31} \citep{Purvis2002:844}. Models of increasing
sophistication have appeared, attempting to re-create this observed
pattern of tree shape observed in nature. For example,
\citet{Heard1996:2141} found that speciation rate variation among
lineages can lead to imbalanced trees. 
\citet{Losos1995:329} found that short ``refractory periods''-- periods
before which a new species can speciate again-- led to more balanced
trees, while \citet{Rogers1996:99} found that very long
refractory periods led to less balanced trees.
\citet{Aldous95,Aldous2001:23} was the first to propose a
(non-evolutionary) model which interpolated between the ERM and the
PDA models. More recently, \citet{Steel2001:91} and
\citet{Pinelis2003:1425} have since developed evolutionary
models which also interpolate.

With these models, one could presumably arrange
parameters to correctly fit the observed pattern of imbalance as
reported by a given statistic. But is that really enough? What if
other aspects of the tree shape, not measured by the statistic, differ
considerably? After all, any single statistic is a one-dimensional
summary of a very complex set of data. One might follow the suggestion
of Agapow and Purvis \citep{Agapow2002:866} and use two different
balance statistics which measure balance in different parts of the
tree, but in this paper we hope to present a more direct approach.

The only proposal made in the literature which has the potential to
encapsulate lots of information about the shape of a tree has been by
\citet{Aldous2001:23}. He suggests first constructing a
scatterplot of the interior nodes, where the $x$ coordinate is the
size of the subclade subtended by that interior node, and the $y$
coordinate is the size of the smaller daughter clade. The proposal is
then to perform nonlinear median regression on
the log-log version of this scatterplot and then use the fitted
function as a descriptor of tree shape. We will call the log-log
scatterplot the ``Aldous scatterplot'' in the following.

There are a number of advantages to this approach. It is very natural
from a statistical viewpoint relative to the other, more ad-hoc,
measures of tree balance. The method has the potential to give quite a
lot of information about tree shape compared to a single summary
statistic. Finally, it allows comparison of trees of different sizes
by superposition of scatterplots, which is a significant advantage.
There is currently no generally accepted method for comparing trees of
different sizes using the standard statistics; this remains a
problematic issue \citep{Mooers1995:379} \citep{stam02a}.

However, there are three disadvantages which may not make Aldous'
proposal as practical as might be hoped. The first is that regression
works best with many points of data, and thus one can only expect his
technique to work with rather large trees. This problem is exacerbated
by the fact that isomorphic subtrees are superimposed on one another
in the scatterplot, further reducing the number of fittable points.
The second is an inherent problem with summarizing a tree as a
scatterplot of this sort. Assume that tree $T$ has two non-isomorphic
subtrees $A$ and $B$ of the same size. Exchanging $A$ and $B$ in $T$
will not change the scatterplot and thus not change any regression
parameters, although the resulting tree may differ significantly in
shape. The third problem is that the resulting output can be hard to
interpret. What does, for example, the $k$th Taylor coefficient of the
fitted function actually signify? Despite these issues, we believe
that this technique is underutilized and may be the technique of
choice when working with large phylogenies.

Overall, it appears that additional methods would be useful for
understanding tree shape. This paper attempts to provide some of these
new methods.

\SBsection{The geometric approach}

\begin{figure}
  \begin{center}
  \includegraphics[angle=0,scale=.75]{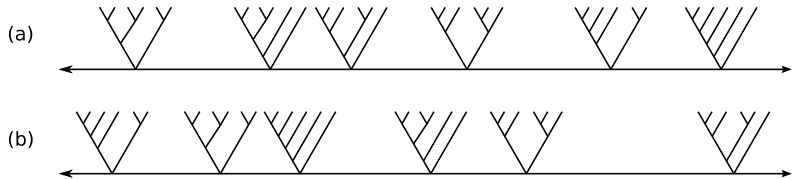}
\end{center}
\caption{Good and bad statistics from the geometric perspective. The
  horizontal axes represent values of hypothetical statistics. In
  figure (a) very different trees are separated, while in figure (b)
  very different trees are close together.}
\label{fig:example_stat}
\end{figure}

The basic philosophy behind the geometric approach is that similar
trees should have similar statistics, and that rather different trees
should have different statistics. This philosophy is summarized in
Figure \ref{fig:example_stat}. All of the trees with six tips are
evaluated by two hypothetical statistics. The top axis shows what one
might consider a good statistic. The maximally balanced tree is on
the far left side, and the completely unbalanced tree is on the far
right. When a subtree is preserved, the statistic tends not to change
too much. The bottom axis shows what might be considered a bad
statistic. The extremes of tree balance are now put together, and two
similar trees are now on the two extremes of the axis.

If we are to apply this sort of intuition on trees, it is necessary to
formalize the notion of similar and different. We do so by
constructing a metric on unlabeled trees.

\SBsubsection{A metric for evolutionary histories}

Here we describe a metric on unlabeled trees which can be applied
directly to compare tree shapes or can be used to guide the selection
of statistics as described below. To begin we state that by ``tree''
we will mean a finite strictly bifurcating rooted tree without leaf
labels or specified edge lengths. We have chosen finite strictly
bifurcating rooted trees, as these correspond most naturally to the
output of models. This paper concerns itself with tree shape rather
than the identity of taxa, thus we consider unlabeled trees. Finally,
our intent in this paper is to understand the combinatorial content of
the tree, and thus we consider trees without specified edge lengths.
The case including edge lengths would be an interesting future
extension of this work, but would require a significant further
development of the methodology.

\begin{figure}
  \begin{center}
  \includegraphics[angle=0,scale=.45]{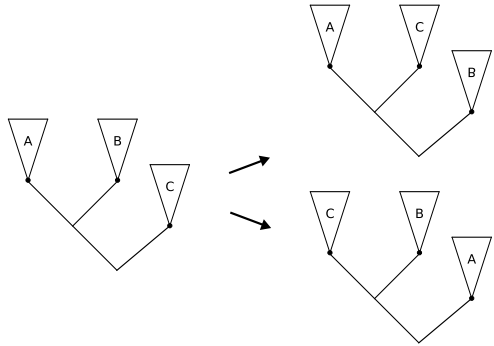}
\end{center}
\caption{A single rooted NNI move.}
\label{fig-nni}
\end{figure}

We recall that a metric $g$ is simply a set of ``distances'' between
pairs of a collection of objects satisfying (i) $g(x,y) = 0$ if and
only if $x=y$, (ii) $g(x,y) = g(y,x)$, (iii) the triangle inequality:
$g(x,y) + g(y,z) \geq g(x,z)$. The metric we consider is simply the
nearest neighbor interchange (NNI) metric on unlabeled trees, depicted
in Figure \ref{fig-nni}. A single NNI ``move'' represents a change of
branching order of a tree to one of two possible configurations. The
unlabeled NNI distance from one tree to another is defined to be the
minimum number of moves necessary to change one tree to the other.
Note that these interchanges have appeared before in 
\citet{Kuhner1995:1421} as proposal draws for their their
Metropolis-Hastings approach to estimating population parameters.

\begin{figure}
  \begin{center}
  \includegraphics[angle=0,scale=.45]{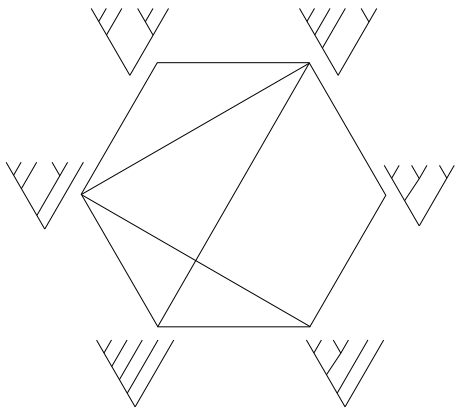}
\end{center}
\caption{Unlabeled tree space equipped with the NNI metric. An edge
  between two trees means that a single NNI move changes one to the
  other.}
\label{fig:tree_space}
\end{figure}

Tree space equipped with the NNI metric is shown in Figure
\ref{fig:tree_space} for trees on 6 leaves. It is a graph which has
connections between any two trees which are a single NNI move apart.
Note that the NNI distance is a special case of the shortest-path
metric on a graph and thus we are justified in calling it a metric.
Also, although the metric is not explicitly model-based, a change of
branching order can be thought of as a change of timing of
diversification events.

Unsurprisingly, computing this metric is NP-complete, as can be seen by
a small modification of a similar proof by \citet{dasgupta}. Their
paper demonstrates that calculating the unrooted NNI distance on
unrooted trees is NP-complete. However, the unrooted NNI moves are
identical to the moves in Figure \ref{fig-nni} when the tree shown in
the diagram is chosen to be anything but the entire tree. Therefore we
can simply root the tree in Figure 4 of their paper on the far left
side of the main linear tree and the proof proceeds as usual.

\SBsubsection{Resolution of Statistics}

In this section we define the notion of ``resolution'' of a tree shape
statistic. Although the formal definition of the resolution is in
terms of the statistical method of multidimensional scaling, we will
first describe how resolution relates to the more common method of
principal component analysis, and then give an intuitive definition of
resolution as a measure of how much a statistic ``spreads out'' the
data. This resolution measure will be applied to various tree shape
statistics below where the underlying data
will be the tree space of a given number of leaves. In this way the
resolution will be our operational definition of performance for tree
shape statistics.

The resolution measure formalizes the intuitive notion that similar
objects should have similar statistics and rather different objects
should have different statistics. For the moment let us consider these
objects to be points in $n$-dimensional space. A natural statistic
which satisfies our criteria is the familiar first principal component
from multivariate statistics. It is some projection of the original
spatial data, so objects which are close together stay close together
after projection. Also, it is the direction along which variance of
the coordinates of the points is maximized, so as much as possible
objects which are far apart stay far apart. In this way we consider
the first principal component to be the best possible statistic for
this collection of points, and will assign it the highest resolution
value.

We can get at the principal component by thinking of it as the
maximization of a certain ``quadratic form.'' In the standard
formulation, the principal components are the eigenvectors of the
covariance matrix constructed from the coordinates of the sample
points. However, it turns out that even if we do not have the actual
coordinates of the points, but rather the distances between them, we
can still construct the covariance matrix. The process goes as follows:
let $H$ be the $n \times n$ ``centering matrix''
\[
H = I - (1/n) J
\]
where $J$ is the matrix with every entry equal to one. The operation
of the centering matrix on a vector subtracts off the average of the
entries of the vector from each component, so the result is a vector
which is perpendicular to the vector of ones. Let $S(A)$ be the
component-wise matrix squaring operation, such that the $ij$ entry of
$S(A)$ is $a_{ij}^2$. Then if $D$ is a ``euclidean distance matrix,''
i.e. a matrix such that the $ij$ entry is the distance between two
points $i$ and $j$ in a euclidean space, then $B = H \, S(D) \, H$
will correspond exactly with the covariance matrix of those same
points calculated in the traditional way \citep{mardiaEA}.

With the covariance matrix now in hand, we can apply the Rayleigh
Quotient theorem, which is a special case of the Courant-Fisher
theorem. It states that the eigenvector corresponding to the largest
eigenvalue of a symmetric matrix maximizes the quadratic form $\x^T M
\x$ 
over all unit-length vectors $x$ \citep{ortega}. Thus in our setting
the first principal component is the unit-norm $\x$ which maximizes
the quadratic form
\begin{equation}
\label{eq:qf1}
R(\x) \ = - \ \x^T H \, S(D) \, H \x.
\end{equation}
Again, the action of left multiplication by $H$ simply subtracts the
average of the components of $\x$. Therefore maximization is certainly
achieved by an $\x$ which has average zero, i.e. is perpendicular to
one. On such $\x$, $H$ clearly has no effect. Therefore we can obtain
first principal component as
\begin{equation}
\label{eq:qf2}
\argmax_{\stack{\norm{\x} = 1}{\x \perp \one}} \ \ \x^T S(D) \x.
\end{equation}
Written out in a slightly longer form this is 
\begin{equation}
\label{eq:qf_intuitive}
\argmax_{\stack{\norm{\x} = 1}{\x \perp \one}} \ \sum_{i,j} -d_{ij}^2 x_i x_j
\end{equation}
This formula has a simple and intuitive explanation. As mentioned
above, in our view a statistic should assign very different values to
objects which are far apart. This equation simply formalizes this
intuition in a nice way: an individual term of the sum in
(\ref{eq:qf_intuitive}) will be maximized if $x_i$ is very negative
and if $x_j$ is very positive. The summation and the distances simply
combine all of these terms together in a weighted fashion such that
$ij$ pairs which are distant carry more weight than ones which are
close. Therefore the more distant objects will tend to be farther
apart in $x$-value, and the closer objects will tend to be closer in
$x$-value.

We will call the quadratic form $R$ of (\ref{eq:qf1}) the
``resolution'' of a statistic, in the sense that a statistic which
differentiates between close and distant objects has a high level of
resolution. As mentioned above, the first principal component
maximizes $R$, and thus its value is an upper limit on the resolution
of a statistic. However, we will see below that some well-known
statistics on tree space achieve resolution nearly that of the first
principal component.

So far we have defined the resolution for data sets of distance
matrices for configuration of points in euclidean space. Although
phrased in a slightly unusual manner, this has led us into the
well-known area of principal component analysis. However, our intent
is to apply this technique to the space of all unlabeled trees with
the NNI metric. The distance matrix corresponding to this space is far
from being a euclidean distance matrix. Is it possible to continue
with the same formalism as in the euclidean setting?

It turns out that we can, and that the procedure is now called metric
multidimensional scaling (MDS) \citep{mardiaEA}. The only difference
is that $D$ is now allowed to be non-euclidean. In essence, when we
substitute a non-euclidean distance matrix into (\ref{eq:qf1}), we
consider the projection of the squared centered matrix onto the cone
of semidefinite matrices. Thus multidimensional scaling performs
principal component analysis on the ``closest'' euclidean distance
matrix to our original matrix in a specific sense \citep{dattorro}.
This operation certainly loses some data, but enough information is
retained to understand the descriptive ability of several statistics.
We visit this issue in the last section.

Note that this is not the first application of MDS to phylogenetic
analysis: \citet{Hillis2005:471} applied it with
interesting results to the space of trees with labeled tips. They used
MDS with the Robinson-Foulds distance metric as a tool for visualization
and analysis of the output of tree reconstruction software. Our intent
and methods differ here, as we are concerned with finding near-optimal
statistics for understanding unlabeled tree space with the NNI metric.

In this section we have defined the resolution as function that allows
us to understand the descriptive ability of some statistic. At this
point we specialize to the case of tree shape statistics on tree space
equipped with the NNI metric. Resolution scores are calculated as
follows: first construct a vector with rows equal to the value of the
statistic on all trees in tree space. Then apply the matrix $H$ to
center the vector; then normalize the vector in the euclidean sense
resulting in a vector $\hat{x}$. The resolution is the value of
$\hat{x}^T S(D) \hat{x}$. We will use this definition to guide
selection of statistics.

\newpage
\SBsection{Results}

In this section the methodology of the previous section is applied to
compare the resolution of tree shape statistics. We will first
evaluate the standard list of statistics \citep{Kirkpatrick1993:1171}
\citep{Agapow2002:866} \citep{felsenstein} according to the above
methodology. Then we search for a best second statistic given the
first, and the best third statistic given a first and second. Our
criterion for performance is high resolution on the whole unlabeled
tree space with the NNI metric as described in the previous section.
The tree space was generated and evaluated by an \texttt{ocaml}
\citep{ocaml} program whose source is available upon request.

We calculated the well-known statistics \nbar\ and \varn\ proposed by
\citet{Sackin1972:225}, \Ic\ proposed by \citet{Colless1982:100}, and
\Bone\ and \Btwo, proposed by \citet{Shao1990:266}. We added
to the list a rarely used statistic \Itwo, invented by
\citet{Mooers1997:31} to provide a measure which weights all nodes
equally. Finally, we implemented the proposal of 
\citet{Aldous2001:23} to perform median regression as described in the
introduction. We fit a quadratic polynomial to the data using median
regression and interpreted the linear and quadratic coefficients as
descriptive statistics which we call \Aone\ and \Atwo. 

We note here that although Aldous' paper did not explicitly specify
how to perform the median regression, we have chosen nonlinear median
regression as described by \citet{Koenker1978:33}.
This method minimizes the sum of the distances of the estimated median
to the data points. Median regression performs much better (as a
maximum-likelihood estimator) than least-squares regression when
errors are non-gaussian, as in our case. It can be easily implemented
using linear programming; in this case it was implemented in 34 lines
of code using an \ocaml\ frontend to the GNU linear programming
package GLPK.

\begin{table}
  \centering
\hspace*{-2cm} 
    \begin{tabular}{cccccccccc}
      $n$ & $\lambda_0$ & \Ic & \nbar & \varn & \Itwo & \Bone & \Btwo & \Aone & \Atwo \\
      \hline
      7 & 7.01 & 6.29 & 6.34 & 6.07 & 5.90 & 6.22 & 6.29 & 2.67 & 2.70 \\
      8 & 21.48 & 19.43 & 19.07 & 18.05 & 17.67 & 18.89 & 19.04 & 5.82 & 6.02 \\
      9 & 48.06 & 43.24 & 43.38 & 41.13 & 39.44 & 42.29 & 42.57 & 7.71 & 8.42 \\
      10 & 125.11 & 116.37 & 115.93 & 110.07 & 103.60 & 111.18 & 111.55 & 31.14 & 33.74 \\
      11 & 299.82 & 283.47 & 282.88 & 268.50 & 249.33 & 269.62 & 269.56 & 84.38 & 89.79 \\
      12 & 755.12 & 714.86 & 714.04 & 676.40 & 626.25 & 676.61 & 672.84 & 224.32 & 241.35 \\
      13 & 1856.88 & 1760.73 & 1760.97 & 1663.67 & 1525.18 & 1661.87 & 1645.81 & 575.67 & 622.98 \\
      14 & 4619.28 & 4387.95 & 4385.72 & 4139.01 & 3779.58 & 4113.12 & 4051.89 & 1458.20 & 1583.53 \\
      15 & 11392.51 & 10819.20 & 10817.17 & 10190.62 & 9241.58 & 10106.57 & 9909.07 & 3788.17 & 4124.96 \\
    \end{tabular}
  \caption{The resolution scores for tree statistics on the NNI
    distance matrix.}
\label{table:first}
\end{table}

\begin{figure}
  \begin{center}
  \includegraphics[angle=-90,scale=.45]{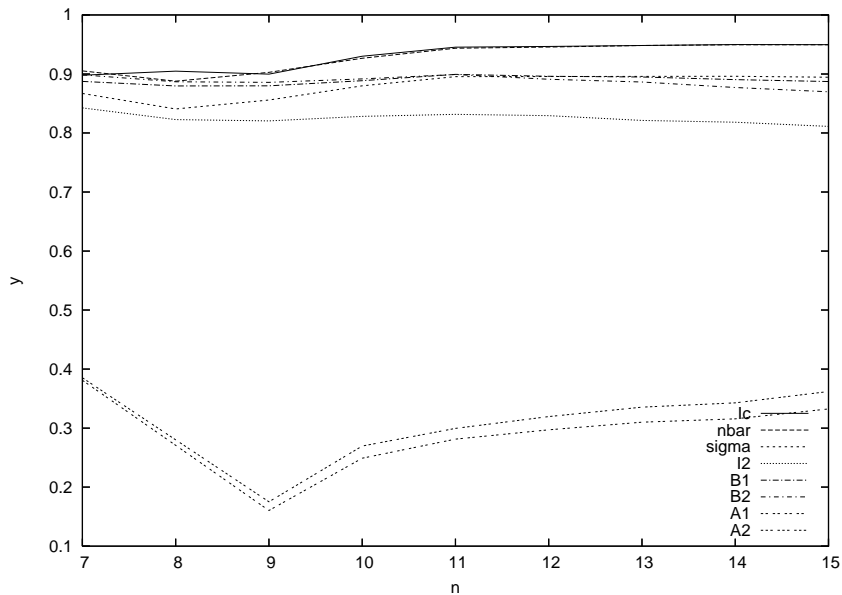}
\end{center}
\caption{Resolution scores divided by the first eigenvalue.}
\label{fig:first}
\end{figure}

The results of this analysis are presented in Table \ref{table:first}
and Figure \ref{fig:first}. First, we find that the resolution of two
statistics, \Ic\ and \nbar, is rather close to the first eigenvalue,
which is the upper limit for the resolution. This is quite remarkable,
in that two statistics which were designed ``by hand'' to measure a
visible aspect of tree shape end up having almost as much resolution
as theoretically possible. The fact that overall tree balance appears
as such an important descriptor justifies in a sense the
disproportionate amount of attention given to it in the tree shape
literature. Another nice fact is that the relative resolution scores
correspond loosely to the power of the statistics as found by
\citet{Agapow2002:866}: \Ic\ and \nbar\ have the most resolution,
followed by \varn\ and \Bone; \Btwo\ has the lowest resolution of the
standard suite of statistics. We report that in this first setting,
\Itwo\ does have substantially lower resolution than the other
statistics, however, we will see that it performs well in later
settings. Finally, it appears that the coefficients of the best-fit
quadratic polynomial on the Aldous scatterplot should not be used as a
first statistic in the simpleminded way presented here on small trees;
it is possible that an alternative formulation would yield better
results.

So far we have only validated that our technique gives results which
do not seem completely out of the ordinary. However, now we can do
something new. Let's say that we choose \Ic\ as our first statistic
and ask the question ``what is the best second number to know about a
tree given that we already know \Ic?'' This question has a
mathematical formulation: we simply project out the \Ic\ component of
the matrix $B$ and repeat the previous process. 

\begin{table}
  \centering
  \begin{tabular}{cccccccc}
n & \nbar & \varn & \Itwo & \Bone & \Btwo & \Aone & \Atwo \\
\hline
7 & 0.15 & 0.03 & 1.89 & 0.75 & 0.53 & 2.68 & 2.74 \\
8 & 0.35 & 0.24 & 5.42 & 1.75 & 1.34 & 6.05 & 6.10 \\
9 & 0.88 & 0.54 & 14.94 & 6.45 & 5.16 & 7.43 & 8.50 \\
10 & 1.85 & 1.77 & 42.47 & 14.55 & 12.37 & 31.72 & 33.76 \\
11 & 4.12 & 5.52 & 110.23 & 40.09 & 35.80 & 85.44 & 89.11 \\
12 & 8.91 & 16.80 & 293.51 & 97.41 & 91.67 & 224.61 & 230.85 \\
13 & 20.06 & 48.81 & 749.81 & 253.42 & 249.96 & 577.10 & 593.12 \\
14 & 44.64 & 139.34 & 1930.63 & 625.33 & 645.74 & 1431.73 & 1449.77 \\
15 & 102.17 & 387.97 & 4883.15 & 1586.90 & 1710.31 & 3657.50 & 3657.96 \\
\end{tabular}
\caption{Resolution scores for tree statistics on the NNI
  distance matrix after projecting out \Ic.}
\label{table:second_a}
\end{table}

The resolution scores of the previously chosen statistics are listed
in Table \ref{table:second_a} with the exception of \Ic, which of
course has resolution zero because we have projected it out. We note
first that \nbar\ has rather small resolution, which is to be expected
because it is highly correlated with \Ic. Comparatively, \Itwo, \Aone, and
\Atwo\ now do better, which means that they measure a different
aspect of tree shape than does \Ic. 

However, it is possible to improve on existing statistics by
explicitly constructing a statistic which measures a different aspect
of tree shape than \Ic. Plotting the principal components of the $B$
matrix suggests that a good second statistic may be the change of
balance from the root to the tips. We have implemented this intuition
in two ways, first as the ``derived statistics'' of a given statistic,
and second as a specific statistic which we call \Qone.

First we describe the construction of the derived statistics of a
given statistic $Y$. Start by making a plot analogous to the Aldous
scatterplot, except now the $x$ axis is the size of the subtree and
$y$ is the value of the statistic $Y$. Now do median regression on
this scatterplot and report the slope of the best-fit line or the
quadratic coefficient of the best-fit quadratic polynomial. Given an
original statistic $Y$ we will call these two derived statistics $Y'$
and $Y''$ in analogy to the first and second derivatives of calculus.
Higher derived statistics are of course possible but will not be
investigated in this paper.

We have designed another statistic, which we call \Qone, which also
attempts to quantify the change of balance from the root to the tips.
The conceptual model for this statistic is the idea that at some time
in the past there may have been a change of evolutionary machinery
such that the balance before that time differs from the balance after
that time. In some sense the procedure tries to find that time and
then compares the balance before and after that time.

The procedure can be described as follows. Begin by assigning to each
internal node a ``local imbalance,'' which quantifies the degree of
imbalance just at that node. If a bifurcating internal node has
subtrees of size $s_l$ and $s_r$, the local imbalance for trees is
\[
\frac{|s_l - s_r|}{s_l + s_r - 2}.
\]
This quantity is similar to the summand in the definition of \Itwo\ by 
\citet{Mooers1997:31}. We set the local imbalance of a
three-node tree to be one at each node. We set the local imbalance of
a two-node tree to be zero unless it is part of a three-node tree.

After local imbalances have been assigned, we iterate up the tree to
find a ``cut'' of the tree into one basal tree and then a collection
of distal trees, which must contain all of the leaves. The cut is
first chosen such that the average local imbalance of the internal
nodes of the distal trees is maximized. Then the first statistic is
computed, which is the average imbalance of the internal nodes of the
distal trees minus the average imbalance of the internal nodes of the
basal tree. This process is repeated to create a second statistic,
except a cut is chosen such that the imbalance of the internal nodes
of the distal trees is minimized. Whichever value is greater in
absolute value is then called \Qone.

We also recall a statistic which has been understood from the
theoretical perspective but which is not in common usage in the tree
shape literature: the number of ``cherries'' of a tree. A ``cherry''
is simply a subtree of two leaves. \citet{mckenzie-steel} have shown
that the distribution of the number of cherries is asymptotically
normal under both the equal rates Markov and the uniform model (see
next section) and have derived the mean and variance for each. 

\begin{table}
  \centering
  \begin{tabular}{ccccccccc}
    n & \Qone & cherries & $\Ic'$ & $\Itwo'$ & $\Bone''$ & $\Btwo''$ &
    \Aone & \Atwo \\
    \hline
    7 & 4.84 & 1.71 & 3.53 & 2.92 & 2.52 & 2.50 & 2.68 & 2.74 \\
    8 & 12.29 & 5.48 & 10.34 & 10.07 & 6.41 & 6.48 & 6.05 & 6.10 \\
    9 & 30.88 & 15.27 & 28.13 & 27.97 & 15.23 & 15.58 & 7.43 & 8.50 \\
    10 & 73.07 & 44.80 & 61.97 & 62.01 & 44.96 & 46.37 & 31.72 & 33.76 \\
    11 & 173.93 & 118.61 & 147.68 & 146.36 & 122.90 & 129.08 & 85.44 & 89.11 \\
    12 & 427.55 & 322.74 & 347.84 & 340.43 & 312.94 & 322.52 & 224.61 & 230.85 \\
    13 & 1024.86 & 833.99 & 871.08 & 868.45 & 798.39 & 823.73 & 577.10 & 593.12 \\
    14 & 2459.67 & 2171.81 & 2127.44 & 2059.13 & 2042.00 & 2101.81 & 1431.73 & 1449.77 \\
    15 & 5972.63 & 5530.14 & 5058.50 & 4873.71 & 5103.33 & 5232.47 & 3657.50 & 3657.96 \\
  \end{tabular}
  \caption{Resolution scores for tree statistics on the NNI
    distance matrix after projecting out \Ic.}
\label{table:second_b}
\end{table}

Table \ref{table:second_b} presents the somewhat surprising results of
the resolution method as applied to the distance matrix after \Ic\ has
been projected out. The best performance is achieved by \Qone, the
somewhat complicated statistic presented above, but close behind is
the number of cherries, perhaps the simplest possible statistic.
Although the performance of the cherry statistic lags behind the above
statistics as a first statistic (see Supplementary Material), it has
remarkably good performance as a second statistic. Similar performance
is achieved by the slightly more complex $\Ic'$. We also report the
values of $B_1''$ and $B_2''$ due to their good performance.

Now assume we choose \Qone\ for our second statistic and look for a
third. As before, we project \Ic\ and \Qone\ out of our matrix and
compare scores. 
\begin{table}
  \centering
  \begin{tabular}{ccccccc}
    $n$ & $\Bone''$ & $\Btwo''$ & $\Qone''$ & $\Ic''$ & \Aone & \Atwo \\
    \hline
    7 & 2.34 & 2.42 & 1.87 & 1.30 & 1.77 & 1.97 \\
    8 & 6.53 & 6.75 & 4.39 & 5.12 & 5.08 & 5.55 \\
    9 & 15.55 & 15.86 & 9.73 & 12.89 & 7.44 & 8.43 \\
    10 & 44.91 & 45.83 & 38.16 & 37.03 & 31.60 & 33.83 \\
    11 & 122.45 & 127.04 & 99.51 & 92.82 & 85.30 & 88.91 \\
    12 & 313.13 & 321.23 & 245.45 & 250.41 & 223.88 & 230.76 \\
    13 & 798.41 & 820.11 & 645.11 & 619.09 & 577.72 & 586.28 \\
    14 & 2040.07 & 2095.10 & 1633.48 & 1524.52 & 1429.47 & 1428.79 \\
    15 & 5104.65 & 5223.00 & 3939.10 & 3822.40 & 3649.16 & 3603.47 \\
  \end{tabular}
  \caption{The resolution scores for tree statistics on the NNI
    distance matrix after projecting out \Ic\ and $\Qone$.}
\label{table:third}
\end{table}
This time it is $\nbar''$ which performs the best. However, we note
that \Aone, \Atwo, and \Itwo\ are not far behind.

In the end, what is the best general-purpose suite of statistics to
use for tree shape description? For a first statistic, the answer is
probably \Ic\ or \nbar. They have high resolution and are simple to
compute. For a second statistic, \Qone\ has the highest resolution but
is somewhat complex; the number of cherries and $\Ic'$ also have good
resolution and simple interpretations. For a third statistic the
statistic with the highest resolution is $\Btwo''$, however if one is
interested in three statistics another good recommendation would be
the triple $(\Ic,\Ic',\Ic'')$ which has satisfactory resolution and
clear interpretation.

\SBsubsection{Example application}

In the introduction, we proposed that ``interpolating'' evolutionary
models could be used to fit any given pattern of overall imbalance. We
argued that this fact motivates the use of multiple tree shape
statistics, as a single statistic may be insufficient to distinguish
between trees generated by the original evolutionary model and a
fitted one. In this section we investigate these matters using
simulations and the results of the previous sections.

The model we have chosen for this example application is Aldous'
``beta-splitting'' model \citep{Aldous95} \citep{Aldous2001:23}. It is
a simple model with a single parameter, $\beta$, which allows
interpolation between the ``comb'' tree ($\beta = -2$) and the
maximally balanced tree ($\beta = \infty$). The ``equal rates Markov''
or ERM tree (i.e. the coalescent tree distribution) emerges when
$\beta = 0$, and the ``proportional to different arrangements'' or PDA
tree (i.e. the distribution on tree shapes induced by a uniform
distribution on labeled trees) appears when $\beta = -1.5$. 

The idea of this model is to recursively split the tips into two
subclades using the beta distribution. More precisely, if we assume
that a clade has $n$ taxa, the probability of the split being between
subclades of size $i$ and $n-i$ is
\[
q_{n,\beta} (i) = C(n;\beta) \frac{\Gamma(\beta+i+1) \Gamma(\beta+n-i+1)}
{\Gamma(i+1) \Gamma(n-i+1)}
\]
where $C(n;\beta)$ is a normalizing constant. This distribution is
equivalent to scattering the taxa on the unit interval and then
splitting with the $B(\beta+1,\beta+1)$ distribution \citep{Aldous95}.

This model is easily adapted to a maximum-likelihood framework. The
likelihood of each tree for a given $\beta$ is the product of the
likelihoods of each split. We consider the likelihood of a
collection of trees to be the product of the likelihoods of each tree.
With a trick from \citep{Aldous95} one can derive a formula for the
$C(n;\beta)$ and then find a $\beta$ which maximizes the log
likelihood of a collection of trees in the standard way.

As an application of the above statistics we investigate the effect of
missing taxa on phylogenetic tree shape using simulation. We will
model the effect on tree shape of a sequencing strategy which is
common in the realm of infectious disease: sequence only those strains
which are significantly different from previously sequenced strains.
We assume that the original tree emerged from an evolutionary process
which has the ERM distribution on trees. We then assume that the edge
lengths are distributed according to a $N(1,.25)$ Gaussian
distribution truncated below zero. Given such a tree with $n$ leaves,
we then recursively delete $k$ taxa in the following manner: find the
pair of taxa which are closest together in terms of tree distance
(including edge length), and randomly delete one of them. We then
perform a maximum-likelihood fit as described above on those trees,
resulting in a $\beta$, and then generate a sample of beta-splitting
trees on $n-k$ leaves using this $\beta$. Which statistics can
distinguish between the original trees and the fitted trees?

We performed this simulation study with a sample size of 500, $n=100$,
and $k=10$. The $\beta$ value fitted to the described deletion process
was $-1.02$, corresponding to a decrease in balance from the $\beta =
0$ original tree. We then compared statistics between 500 of the ``fitted''
beta-splitting trees and the original trees with deleted taxa. The
trees were then evaluated with the two-tailed Wilcoxson rank sum test
to find statistical power of each statistic to differentiate between
the two distributions. The results of this analysis are in Table
\ref{table:example}.

Remarkably, the statistical power for this scenario corresponds with
the resolution of these statistics when \Ic\ has been projected out.
This makes some sense because when we fit a tree to the beta-splitting
model, we are primarily fitting the overall balance of the trees. We
recall that the four statistics with highest resolution after
projection were \Qone, the number of cherries, $\Ic'$, and \Itwo.
Three out of four of these statistics are also the most powerful for
our example application. Although this is an indicative
correspondence, one reason it is not perfect is that the resolution
scores trees based on overall descriptive ability and here we consider
statistical power to differentiate between two specific models. For
example, considering that cherries tend to be eliminated by the
described taxon deleting process, it is not surprising that the number
of cherries would have such high statistical power in this example
application. We have also included the statistics \Aone\ and \Atwo\ in
Table \ref{table:example} because they performed reasonably well; this
corresponds with their good resolution after projecting out \Ic\ as
shown in Table \ref{table:second_a}. It is not surprising that these
statistics perform better on relatively large trees. Finally, as might
be expected for a situation in which we have fitted the overall
balance of a tree to the model, the statistic \Ic\ has essentially no
power to distinguish between the two models.

\begin{table}
\centering
\begin{tabular}{cccccccc}
& \Ic & cherries & \Itwo & \Qone & $\Ic'$ & \Aone & \Atwo \\
\hline
NM & 0.077 & 30 & 0.47 & 0.24 & 0.015 & 0.62 & 0.056 \\
DM & 0.076 & 29 & 0.49 & 0.27 & 0.019 & 0.51 & 0.089 \\
$p$ & 0.16 & 7.6e-32 & 5.1e-13 & 1.8e-07 & 4.6e-07 & 4.4e-06 & 1.1e-06 \\
\end{tabular}
\caption{Comparison of the scores for various statistics when
  applied to trees from two different models. ``NM'' signifies the
  median score of the statistic when applied to a sample of ERM
  trees of size 90; ``DM'' signifies the median when applied to a
  sample of beta-splitting trees with leaves deleted as described in
  the text. The last line shows the $p$-value for the two-sided Wilcoxson
  rank-sum test.}
\label{table:example}
\end{table}

We argue that this simple simulation exercise further demonstrates
that the resolution measure can help guide the selection of good
general-purpose tree shape statistics. Although these statistics were
chosen on purely geometric grounds, they were also the most powerful
for this somewhat arbitrary model.

\SBsection{Extensions}

There are a number of limitations to this methodology which point the
way for future development. The first is that this application of the
MDS technique was to a specific model of tree space, namely that with
the unlabeled NNI distance. It is possible that this is not a good
choice. However, if another model is found which seems more
appropriate, that can be easily brought into the general framework
presented here and derive analogous results. Another angle of this
problem is that the resolution parameter described implicitly takes
the uniform distribution on trees. That is to say, trees which are
never seen in models or from data carry equal weight in the resolution
measure as trees which are common. This could decrease the utility of
the resolution measure, especially when considering large trees.
However, in the author's opinion there is no clear choice of
distribution. In fact, the main purpose of tree shape theory is to
think about what sorts of distributions are appropriate for tree
shape. If a clear alternative distribution is found, some
modifications will have to be made to the methodology to incorporate
this information.

Second, this methodology offers nothing to the debate of how to
compare the shape of trees of different size. This is a very
fundamental problem which may be more philosophical than technical:
what does it actually mean to say that a tree of one size has a
similar shape to one of a different size? A common response in the
literature \citep{Mooers1995:379} \citep{stam02a} is to compare in one
way or another the shape of a given tree to a sample of trees from a
fixed distribution; knowing the distribution of the statistic as for
the number of cherries \citep{mckenzie-steel} makes this an attractive
option for some statistics. However, if we wish to have a descriptive
theory independent of perhaps over-simple models, some other method
will have to be found. This is clearly an interesting avenue for
future research.

Third, because the number of unlabeled binary trees is very large,
asymptotically $O(c^n n^{-3/2})$ \citep{harding} \citep{semple-steel},
we have had to limit ourselves to moderately small trees. This may
skew the analysis in that statistics which perform poorly for small
trees may perform quite well for large trees; an example case might be
Aldous' descriptors of tree shape. One response to this objection is
that Figure \ref{fig:first} shows a certain level of stability as $n$
increases: statistics which are good for smaller $n$ appear to be good
for larger $n$ as well. As our understanding of this NNI tree space is
very limited, we cannot prove any statement of this type at this time.
Furthermore, although increasingly large trees are now available, the
analysis of trees of intermediate size is still a challenge and at
worst the above methodology is applicable to that case. However, we do
consider this to be a problem for future research.

Fourth, multidimensional scaling with non-euclidean data always loses
some information. This results from the fact that the analysis is
actually performed on a projection of the original distance matrix. As
mentioned, the NNI tree space is certainly non-euclidean: even in the
innocuous-looking case of $n=6$ (see Figure~\ref{fig:tree_space}) some
distortion results from a euclidean projection. The subject of how
much information is lost from this projection is very interesting but
requires a separate treatment. We will address these issues in a
future article.

Fifth, edgelength information is conspicuously absent in tree shape
analysis. Typically information about timing of speciation (or other
branching) events is analyzed in a completely different manner, as a
lineages-through-time plot, which is then used to estimate speciation
and extinction rates with maximum likelihood \citep{neeEA94a}. Clearly
any analysis of this sort eliminates topological information which may
aid in choosing an evolutionary model. The tree shape literature has
already shown that the standard birth-death process where each leaf is
equally likely to split or be eliminated does not construct trees
which seem to reflect the imbalance seen in nature; nevertheless this
assumption is implicit in Nee et. al.'s analysis. More work is needed
to integrate the tree shape and timing literature.

Finally, we come to a limitation which is fundamental to any
discussion of trees: with very few exceptions, trees are not actual
data. They are almost certainly flawed reconstructions of historical
events. A common response to this problem by coalescent theorists
trying to estimate evolutionary parameters is to simply ``integrate
out'' the history by performing MCMC iteration over all possible
histories \citep{Kuhner1995:1421}. However, we believe that there is a
signal in tree shape that stands out from the noise and which can
guide us in selection of evolutionary models. We also note that tree
shape has a role in understanding potential problems and biases of
tree reconstruction methods.

In summary, we have developed a new method for evaluating tree shape
statistics, which we call the ``resolution'' of a statistic. This
method formalizes the intuition that a good statistic takes on similar
values for similar trees and different values for rather different
trees. It has the advantage that it can help choose a $k$th statistic
given that $k-1$ other statistics are already known; this opens up the
possibility of finding a useful suite of statistics to describe a
tree. We then use the method to make specific recommendations for such
a suite of three statistics. Finally, we compare the results of the
geometric analysis to two model-based tree distributions and find that
statistics with good resolution were also the ones which had high
power to distinguish the two distributions. We hope that these
statistics and methodology will prove useful for scientists engaged in
the fascinating questions emerging from macroevolution and
phylogenetic reconstruction. We suggest that this paper represents a
small step in an area which will continue to pose interesting
questions for years to come.

\SBsubsubsection{Acknowledgments}

\begin{footnotesize}
  The author would like to thank Akira Sasaki for asking him the
  question ``what is a good way to numerically describe the shape of a
  tree?'' two years ago, as well as David Aldous, Steve Evans, Joseph
  Felsenstein, Susan Holmes, Arne Mooers, Montgomery Slatkin and John
  Wakeley for stimulating discussion and valuable comments. F.A.M. was
  supported by a Graduate Research Fellowship from the National
  Science Foundation.
\end{footnotesize}

\newpage
\bibliographystyle{plainnat}
\bibliography{/home/matsen/papers/bibtex_entries,good}

\begin{thebibliography}{33}
\expandafter\ifx\csname natexlab\endcsname\relax\def\natexlab#1{#1}\fi
\expandafter\ifx\csname url\endcsname\relax
  \def\url#1{{\tt #1}}\fi

\bibitem[Agapow and Purvis(2002)]{Agapow2002:866}
P.M. Agapow and A.~Purvis.
\newblock Power of eight tree shape statistics to detect nonrandom
  diversification: A comparison by simulation of two models of cladogenesis.
\newblock {\em Syst. Biol.}, 51\penalty0 (6):\penalty0 866--872, 2002.

\bibitem[Aldous(1995)]{Aldous95}
D.J. Aldous.
\newblock Probability distributions on cladograms.
\newblock In David Aldous and R.~Pemantle, editors, {\em Random Discrete
  Structures}, pages 1--18. Springer, Berlin, 1995.

\bibitem[Aldous(2001)]{Aldous2001:23}
D.J. Aldous.
\newblock Stochastic models and descriptive statistics for phylogenetic trees,
  from yule to today.
\newblock {\em Stat. Sci.}, 16\penalty0 (1):\penalty0 23--34, 2001.

\bibitem[Chailloux et~al.(2000)Chailloux, Manoury, and Pagano]{ocaml}
E.~Chailloux, P.~Manoury, and B.~Pagano.
\newblock {\em D\'{e}veloppement d'applications avec Objective CAML}.
\newblock O'Reilly, Sebastopol, CA, 2000.
\newblock English translation available at
  http://caml.inria.fr/pub/docs/oreilly-book/.

\bibitem[Colless(1982)]{Colless1982:100}
D.H. Colless.
\newblock Phylogenetics: the theory and practice of phylogenetic systematics.
\newblock {\em SYST ZOOL}, 31\penalty0 (1):\penalty0 100--104, 1982.

\bibitem[DasGupta and et. al.(2000)]{dasgupta}
B.~DasGupta and et. al.
\newblock On computing the nearest neighbor interchange distance.
\newblock In D.~Z. Du, P~M. Pardalos, and J.~Wang, editors, {\em Proceedings of
  the DIMACS Workshop on Discrete Problems with Medical Applications},
  volume~55 of {\em DIMACS Series in Discrete Mathematics and Theoretical
  Computer Science}, pages 125--143. American Mathematical Society, 2000.

\bibitem[Dattorro(2005)]{dattorro}
J.~Dattorro.
\newblock {\em Convex Optimization and Euclidean Distance Geometry}.
\newblock Meboo Publishing, USA, 2005.
\newblock available at http://www.stanford.edu/~dattorro/.

\bibitem[Felsenstein(2004)]{felsenstein}
J.~Felsenstein.
\newblock {\em Inferring Phylogenies}.
\newblock Sinauer Press, Sunderland, MA, 2004.

\bibitem[Guyer and Slowinski(1991)]{Guyer1991:340}
C.~Guyer and J.B. Slowinski.
\newblock Comparisons of observed phylogenetic topologies with null
  expectations among 3 monophyletic lineages.
\newblock {\em Evolution}, 45\penalty0 (2):\penalty0 340--350, 1991.

\bibitem[Guyer and Slowinski(1993)]{Guyer1993:253}
C.~Guyer and J.B. Slowinski.
\newblock Adaptive radiation and the topology of large phylogenies.
\newblock {\em Evolution}, 47\penalty0 (1):\penalty0 253--263, 1993.

\bibitem[Harding(1971)]{harding}
E.~F. Harding.
\newblock The probabilities of rooted tree-shapes generated by random
  bifurcation.
\newblock {\em Advances in Appl. Probability}, 3:\penalty0 44--77, 1971.
\newblock ISSN 0001-8678.

\bibitem[Heard(1996)]{Heard1996:2141}
S.B. Heard.
\newblock Patterns in phylogenetic tree balance with variable and evolving
  speciation rates.
\newblock {\em Evolution}, 50\penalty0 (6):\penalty0 2141--2148, 1996.

\bibitem[Hillis et~al.(2005)Hillis, Heath, and john]{Hillis2005:471}
D.M. Hillis, T.A. Heath, and K.~St john.
\newblock Analysis and visualization of tree space.
\newblock {\em Syst. Biol.}, 54\penalty0 (3):\penalty0 471--482, 2005.

\bibitem[Kirkpatrick and Slatkin(1993)]{Kirkpatrick1993:1171}
M.~Kirkpatrick and M.~Slatkin.
\newblock Searching for evolutionary patterns in the shape of a phylogenetic
  tree.
\newblock {\em Evolution}, 47\penalty0 (4):\penalty0 1171--1181, 1993.

\bibitem[Koenker and Bassett(1978)]{Koenker1978:33}
R.~Koenker and G.~Bassett.
\newblock Regression quantiles.
\newblock {\em Econometrica}, 46\penalty0 (1):\penalty0 33--50, 1978.

\bibitem[Kuhner et~al.(1995)Kuhner, Yamato, and Felsenstein]{Kuhner1995:1421}
M.K. Kuhner, J.~Yamato, and J.~Felsenstein.
\newblock Estimating effective population-size and mutation-rate from sequence
  data using metropolis-hastings sampling.
\newblock {\em Genetics}, 140\penalty0 (4):\penalty0 1421--1430, 1995.

\bibitem[Losos and Adler(1995)]{Losos1995:329}
J.B. Losos and F.R. Adler.
\newblock Stumped by trees - a generalized null model for patterns of
  organismal diversity.
\newblock {\em Am. Nat.}, 145\penalty0 (3):\penalty0 329--342, 1995.

\bibitem[Mardia et~al.(1979)Mardia, Kent, and Bibby]{mardiaEA}
K.V. Mardia, J.T. Kent, and J.M. Bibby.
\newblock {\em Multivariate Analysis}.
\newblock Academic Press, New York, 1979.

\bibitem[McKenzie and Steel(2000)]{mckenzie-steel}
Andy McKenzie and Mike Steel.
\newblock Distributions of cherries for two models of trees.
\newblock {\em Math. Biosci.}, 164\penalty0 (1):\penalty0 81--92, 2000.
\newblock ISSN 0025-5564.

\bibitem[Mooers(1995)]{Mooers1995:379}
A.O. Mooers.
\newblock Tree balance and tree completeness.
\newblock {\em Evolution}, 49\penalty0 (2):\penalty0 379--384, 1995.

\bibitem[Mooers and Heard(1997)]{Mooers1997:31}
A.O. Mooers and S.B. Heard.
\newblock Evolutionary process from phylogenetic tree shape.
\newblock {\em Q. Rev. Biol.}, 72\penalty0 (1):\penalty0 31--54, 1997.

\bibitem[Nee et~al.(1994)Nee, May, and Harvey]{neeEA94a}
S.~Nee, R.~M. May, and P.~H. Harvey.
\newblock The reconstructed evolutionary process.
\newblock {\em Philos Trans R Soc Lond B Biol Sci}, 344\penalty0
  (1309):\penalty0 305--11, May 1994.

\bibitem[Ortega(1987)]{ortega}
J.~M. Ortega.
\newblock {\em Matrix Theory: a Second Course}.
\newblock Plenum Press, New York, 1987.

\bibitem[Pinelis(2003)]{Pinelis2003:1425}
I.~Pinelis.
\newblock Evolutionary models of phylogenetic trees.
\newblock 270\penalty0 (1522):\penalty0 1425--1431, 2003.

\bibitem[Purvis and Agapow(2002)]{Purvis2002:844}
A.~Purvis and P.M. Agapow.
\newblock Phylogeny imbalance: Taxonomic level matters.
\newblock {\em Syst. Biol.}, 51\penalty0 (6):\penalty0 844--854, 2002.

\bibitem[Rogers(1996)]{Rogers1996:99}
J.S. Rogers.
\newblock Central moments and probability distributions of three measures of
  phylogenetic tree imbalance.
\newblock {\em Syst. Biol.}, 45\penalty0 (1):\penalty0 99--110, 1996.

\bibitem[Sackin(1972)]{Sackin1972:225}
M.J. Sackin.
\newblock Good and bad phenograms.
\newblock {\em SYST ZOOL}, 21\penalty0 (2):\penalty0 225--226, 1972.

\bibitem[Savage(1983)]{Savage1983:225}
H.M. Savage.
\newblock The shape of evolution - systematic tree topology.
\newblock {\em Biol. J. Linnean Soc.}, 20\penalty0 (3):\penalty0 225--244,
  1983.

\bibitem[Semple and Steel(2003)]{semple-steel}
C.~Semple and M.~Steel.
\newblock {\em Phylogenetics}.
\newblock Oxford University Press, New York, 2003.

\bibitem[Shao and Sokal(1990)]{Shao1990:266}
K.T. Shao and R.R. Sokal.
\newblock Tree balance.
\newblock {\em SYST ZOOL}, 39\penalty0 (3):\penalty0 266--276, 1990.

\bibitem[Simon and Page(2005)]{simon-page}
Chris Simon and Rod Page.
\newblock The past and future of systematic biology.
\newblock {\em Systematic Biology}, 54\penalty0 (1):\penalty0 1--3, 2005.

\bibitem[Stam(2002)]{stam02a}
Ed~Stam.
\newblock Does imbalance in phylogenies reflect only bias?
\newblock {\em Evolution Int J Org Evolution}, 56\penalty0 (6):\penalty0
  1292--5, Jun 2002.

\bibitem[Steel and Mckenzie(2001)]{Steel2001:91}
M.~Steel and A.~Mckenzie.
\newblock Properties of phylogenetic trees generated by yule-type speciation
  models.
\newblock {\em Math. Biosci.}, 170\penalty0 (1):\penalty0 91--112, 2001.

\end{thebibliography}

\SBsection{Supplementary Material} 

Here I will present tables of all of the statistics, not just the ones
with high resolution values.

\end{document}